\documentstyle[aas2pp4]{article}       

\lefthead{Menci n\&vCavaliere}
\righthead {Groups and Clusters of Galaxies}

\begin{document}
\title{The History of Cosmic Baryons: X-ray emission vs. Star Formation Rate}
\author{N. Menci}
\affil{Osservatorio Astronomico di Roma,
 via Osservatorio, 00040 Monteporzio, Italy}
\centerline{\&} 
\author{A. Cavaliere}
\affil{Astrofisica, Dipartimento Fisica, II Universit\`a di Roma,\\
via Ricerca Scientifica 1, 00133 Roma, Italy}

\begin{abstract}
We relate the star forma<tion from {\it cold} baryons {\it condensing}
in virialized structures to the X-ray properties of the associated
diffuse, {\it hot} baryonic component. Our computations use the
standard ``semi-analytic'' models to include and
connect three sectors of the complex astrophysics involved: first,
the formation of dark matter halos through accretion and merging,
after the standard hierarchical clustering; second, the star formation
governed, after the current ``recipes'', by radiative cooling and by
feedback of the supernova energy into the hot baryonic component;
third, and novel, the hydro- and thermodynamics of the hot phase,
rendered with our Punctuated Equilibria model.  So we relate the X-ray
observables concerning the intra-cluster medium (namely, the
luminosity - temperature relation, the luminosity functions, the
source counts) to the thermal energy of the gas pre-heated and
expelled by supernovae following star formation, and then accreted
during the subsequent merging events. 

Our main results are as follows. At fluxes fainter than $F_X\approx
10^{-15}$ erg/cm$^2\,$ s the X-ray counts of extended extragalactic
sources (as well as the faint end of the luminosity function, their
contribution to the soft X-ray background, and the $L_X-T$ correlation
at the group scales) increase considerably when the star formation
rate is enhanced for $z > 1$ as indicated by growing optical/infrared
evidence. Specifically, the counts in the range 0.5-2
keV are increased by factors $\sim $ 4 when the the feedback is decreased 
and star formation is enhanced as to yield a flat shape of the star 
formation rate for $2<z<4$.

Such faint fluxes are well within the reach of next generation X-ray
observatories like AXAF and XMM. So very faint X-ray counts will soon
constitute a new means of gaining information about the stellar
processes (formation, and supernova feedback) at $z>2$, and a new
way to advance the understanding of the galaxy formation. 
\newline 
 \end{abstract}
{\bf key words:} galaxies: clusters: general -- intergalactic 
medium -- X-rays: galaxies -- galaxies: formation. 

\newpage
\section{INTRODUCTION}

The history and the fate of cosmic baryons constitute 
an outstanding problem in cosmology. It is matter of current debate 
their time-shifting partition among the stars and the diffuse cold and 
hot components, the latter being heated by stellar activity or at the 
expenses of the gravitational energy in the dark matter (DM) 
condensations. 

Two lines of evidence enter the argument: one  comes from the canonical bands, 
optical-UV and IR; the other 
will come -- we argue -- from the X-ray band. 

The first assessments from O-UV data of the global star formation rate
(SFR) suggested a peak at $z\approx 1.5$ with a sharp decline out to
$z\approx 4$ (Madau et al. 1996; Connolly et al. 1998); in such a
picture, less than 20 \% of the stars would have been formed at
$z>2$. Based on the hierarchical cold dark matter (CDM) models of
structure formation, which predict 
the galaxy assembly to be a gradual process proceeding through merging and
accretion events, 
such history has been modeled (see White \& Frenk 1991;
Cole et al. 1994; Kauffmann, White \& Guiderdoni 1993; 
Baugh et al. 1998) by strongly coupling 
the star formation (SF) inside the halos to 
their dynamical growth; in fact, the SFRs have been assumed to scale
up considerably with the DM halo masses.

However, such scenario has been seriously challenged by more recent
observations (Pettini et al. 1997; Dickinson 1998; Meurer et
al. 1997) indicating that the apparent decline for $z>2$ was seriously
affected by dust extinction; the underlying SFR may be increased by
factors 3-15, up to yield a flat plateau for $z>2$ (Madau, Pozzetti \&
Dickinson 1998).  This picture is confirmed by the large statistics
obtained in denser fields by Steidel et al. (1998).  Supporting
indipendent evidence comes from the still scanty IR data obtained with SCUBA
(see Smail, Ivison \& Blain 1997; Hughes et al. 1998; Barger et
al. 1999).

But the issue is still at stake. In fact, a flat SFR might easily
lead to an overproduction of the IR background, and of the metal
adundance in high-$z$ absorbers (see Ellis 1998). Moreover, it
is not clear how the optical fields sampled so far 
should be weighted to yield a reliable estimate of the average SFR.

Luckily, there is another angle to the baryon history. This is provided by
the hot baryonic phase which is related to the SF and can
be observed directly in X-rays as intra-cluster medium (ICM). A large
amount of baryons corresponding to $\Omega_b/\Omega\approx 0.1$ is
observed at $z\approx 0$ in galaxy groups and clusters on scales
around $R\sim 1$ Mpc (White et al. 1993; White \& Fabian 1995; see
also Fukugita, Hogan \& Peebles 1998).  Such baryons are heated up to
virial temperatures $kT\sim GM/10\,R$ in the gravitational potential
wells corresponding to masses $M\approx 10^{13} -
10^{15}\,M_{\odot}$; at densities $n\approx 10^{-3}$ cm$^{-3}$ they
emit by thermal bremsstrahlung copious X-rays with luminosities
$L\propto R^3\,n^2\,T^{1/2}\approx 10^{44}$ erg/s. But the ICM 
is not secluded; indeed, the shape of the $L_X-T$ correlation,
which is observed to be bent from $L\propto T^2$ in very rich clusters
(Allen \& Fabian 1998) to a much steeper shape in groups (Ponman et
al. 1996), indicates that the temperature of the external, infalling
gas plays a key role in determining the X-ray properties of small groups and
poor clusters (Cavaliere, Menci \& Tozzi 1997, 1998; Balogh, Babul \&
Patton 1998). Such external temperature, in turn, may be 
set by supernova
feedback to values exceeding the virial value in small merging lumps, so a 
connection is envisaged between the stellar processes and the astrophysics 
of the X-ray emitting ICM.

This connection will provide a relationship between the O, UV or IR
observations and the X-ray data, which is the scope of this paper.  In
Cavaliere, Menci \& Tozzi (1997, 1998a, 1998b hereafter CMT97, CMT98a,
CMT98b) we have already developed a semi-analytical approach to the
X-ray emission, to deal with the DM merging histories, with the
associated infall of external preheated gas, and with the resulting
shock compression, all concurring to estabilish a sequence of
approximate hydrostatic equilibria of the ICM (Punctuated Equilibria
model). Our predicted, bent $L_X-T$ correlation is in eccellent
agreement with the observations; the parameter of the model is the
initial temperature of external gas, which was assumed at a constant
value around 0.5 keV adopted from simple estimates in the literature. 

Such a temperature is of key importance as it happens to be close to
the virial temperatures in typical groups, and may be surmised to rise
gradually during structure formation. In addition, the amount of gas
expelled by supernova feedback is essential in determining the amount of 
baryons filling of shallow potential wells, and hence the X-ray
luminosity of small groups.  So here we set out to compute these
processes in full, based on the detailed recipes currently implemented
in the semi-analytical models for star and galaxy formation (SAM
hereafter, see Kauffmann, White \& Guiderdoni 1993; Cole et al. 1994;
Baugh et al. 1998; Somerville \& Primack 1998) in
hierachically evolving DM halos.  This allows us to connect in a {\it
single} framework the hierarchical clustering, the astrophysics of
star forming baryons, and the thermo- and hydrodynamics of the X-ray
emitting plasma.  In particular, we will show how forthcoming
observations with AXAF can X-ray the SF history, so complementing the
O-IR information.

\section{THE MODEL}

Toward our scope we need to 
relate both the SF and the astrophysics of the ICM 
to the gravitationally dominant DM content of the halos.  
After the hierarchical clustering paradigm, 
 the latter form due to the gravitational 
instability of initial random density fluctuations, and then evolve through 
stochastic merging of smaller or sometimes comparable 
units into larger structures. The resulting 
statistics is well estabilished  in the form of the 
extended Press \& Schechter theory (EPST, see Bond et al. 1991;
White \& Frenk 1991; Bower 1991; Lacey \& Cole 1993).

As to the process of SF, we adopt an {\it analytical} 
rendition of the SAM approach. In the above papers 
the  baryons gravitationally bound to a DM halo, after 
being heated to its virial 
temperature, are assumed to be exchanged among three phases: 
the {\it cold} phase produced by radiative cooling;  
the resulting {\it condensed} phase into stars; the {\it hot} phase 
constituted by gas preheated by supernova explosions and further raised to 
the virial temperature of the potential wells. 
The model includes coalescence of baryons following halo merging, and 
the luminosity-color evolution due to the rise and fall of successive star 
generations.

We include the  X-ray emission in the very same framework, using 
the model proposed by CMT97: at each time step 
the X-ray luminosity is computed from the hot phase of the ICM, and from 
 the boundary conditions relating this to the external gas infalling in each 
 merging event. As shown in CMT97-98, such boundary conditions are essential 
 in providing a realistic model for the X-ray emission from clusters and 
groups. In fact, a set of equilibrium states exists for the ICM in a given 
halo, each corresponding to a different merging history and hence 
to a different boundary condition. 
The average X-ray luminosity corresponding to 
a given mass $M$ is then obtained by convolving over all possible 
merging histories the emission corresponding to each equilibrium state.

Because of the added complexity to the standard sense, and of the
related increase in computing time, we choose to express the merging
histories analytically with the EPST rather than using the equivalent
Montec Carlo simulations.  Below we describe the basic ingredients of
our model.

Whenever necessary, and untill otherwise specified, the redshift $z$ 
will be associated with the epoch $t$ on adopting a critical Universe with 
$H_o=50$ km/s Mpc. 

\subsection{Gas Cooling, Feedback and Star Formation Rates}

Our recipes to relate the SFR to the dynamics of DM halos are basically
taken from Cole et al. (1994) and are summarized as follows. 

a) For each DM mass $M$ (or circular velocity $v_c \propto M^{1/3}$), 
the baryonic
content is divided into a cold gas phase [apt to radiatively
cool within the halo] with mass $m_c$, into a star phase with mass
$m_*$, and into the complementary hot phase with mass $m_h$.  The
initial gas content associated to all the halos sums up to the
universal baryonic density $\Omega_b$. The initial stellar content at
$z_i\sim 10$ is taken to be nil.

b) The mass of the cold phase is increased from inside out by cooling
processes: at each time step $\Delta t$, we have then $\Delta
m_c(v_c)= 4\pi\,r_{cool}^2\,\rho_g\,\Delta r_{cool}$, where gas density 
$\rho_g$ and the dependencies
of the cooling radius $r_{cool}$ on $v_c$ and on $t$ are given, e.g.,
by Somerville \& Primack (1999). We adopt the cooling function used 
by Cole et al. (1994), which was computed assuming a primordial mixture of 
77 \% hydrogen and 23 \% helium. 

c) Then it is computed the amount of cold gas which goes into stars,
and the corresponding mass is transferred from the cold phase to the
star phase.  Such amount $\Delta m_*(v_c,t) = \dot m_*\,\Delta t$ is
regulated by the SFR, namely $\dot
m_*(v_c,t)=m_c/\tau_*$.  The timescale $\tau_*$ is parametrized in the form
$\tau_*(v_c)=\tau_*^0\,(v_c/300~{\rm km/s})^{\alpha_*}$; the cold mass 
$m_c$ is updated after each time step by subtracting not only the mass of stars 
which have formed but also the cold gas expelled given below. 

d) A final transfer is due to supernova feedback from the cold back to
the hot phase; this involves the amount $\Delta m_h = \beta(v_c) \,\Delta
m_*(v_c,t)$, where the feedback per unit star mass is given by 
$\beta = (v_c/v_h)^{-\alpha_h}$, with the parameters $v_h$ and $\alpha_h$.

For each circular velocity $v_c$, we compute the further variations of
the masses of the gas components and of the star content due to
merging of DM halos (treated in detail in \S 2.3), possibly followed
by galaxy coalescence. The latter is included on considering the
probability that the galaxy coalescence time (parametrized as in Cole
et al. 1994) is shorter than the halo survival time (depending on $v_c$), 
and by averaging over
all possible halo merging histories (details are given in Poli et
al. 1999). The values of the parameters $\tau_*^0$, $\alpha_*$, $v_h$
and $\alpha_h$ are discussed in \S 3.

\subsection{The X-ray Emitting ICM}

At the time $t$, the X-ray luminosity due to bremsstrahlung emission 
by the hot baryons $m_h$ in a
halo of mass $M$ is given by (CMT98b)
\begin{equation}
L_X=A\bigg({m_h\over M}\bigg)^2G^2(M,M')\,T_v^2(M)\,\rho^{1/2}\,I(M,t).
\end{equation}
Here $T_v(M)=4\,(M/10^{15}\,M_{\odot})^{2/3}\,(1+z)$ keV is the virial
temperature corresponding to $M$; the average DM density inside the
cluster is $\rho \propto (1+z)^3$; the normalization constant $A$ will
be adjusted so as to match the height of the observed local $L_X-T$ relation at
$T=4$ keV; the shape factor $I$ describes the internal ICM
distribution. Finally, $G$ is the density jump (ratio) across the shock
induced at the cluster boundary, at around the virial radius, 
by the infalling gas; this will depend
not only on the cluster mass $M$, but also on the mass $M'$ of the
infalling clumps. 

The contribution of emission lines has been included using the standard and 
public Raymond-Smith code. 

We stress why the $L_X-T$ dependence differs from the self-similar
power-law $L\propto T_v^2$. To a small degree this is due to the shape
factor $I(M,t)$, a slowly varying function of $M$ and $t$ of the
detailed form given by CMT98b.  This includes the integration
over the cluster volume of the internal density run (normalized to the
value at the virial radius); the latter is provided by the hydrostatic
equilibrium once the gravitational potential $\phi (M)$ associated to
$M$ is given (we shall adopt the form given by Navarro,
Frenk \& White 1997).

Much more important is the strong dependence of $G^2$ on the halo 
merging histories; in fact, the (squared) compression
ratio $G^2$ at the shock connects the ICM with the infalling gas associated 
with the merging partner of mass $M'$.
Of this, the fraction $f_*$ hovers around the clump $M'$ having being 
expelled and heated to a temperature $T_*(M')$ by supernova feedback;
the complementary fraction $1-f_*$ is still contained inside the
virial radius of $M'$ at the virial temperature $T_v(M')$.  During a
merging event, both components fall into our well of mass 
$M$, and  the compounded $G^2$ entering the luminosity (1) reads
\begin{equation}
G^2(M,M')=f_*\,g^2[T_*(M')]+(1-f_*)g^2[T_v(M')].
\end{equation}
Here $g(T)$ denotes the shock compression factor for the inclusion of each 
gas fraction into a 
cluster or group, given in CMT98a,b to read 
\begin{equation}
g(T)=2\,\Big(1-{T\over T_2}\Big)+\Big[4\, 
\Big(1-{T\over T_2}\Big)^2 + {T\over T_2}\Big]^{1/2}~; 
\end{equation} 
the expression for $T_2$ is given by the same authors, and for strong shocks 
reads simply
$$
k\,T_2(M)\approx {{\mu m_H \phi(M)}/3} + 3k\,T/2~, 
$$
where $m_H$ is the proton mass. 
As anticipated above eq. (2), the argument $T$ in eq. (3) 
takes one of the following two values: 
either the virial temperature (with the 
coefficient given by Metzler \& Evrard 1997) 
\begin{equation}
T_v(M')=4\,(M'/10^{15}\,M_{\odot})^{2/3}\,(1+z)~~ {\rm keV}~  
\end{equation}
for the $1-f_*$ baryons retained in the lump $M'$; or
the temperature of the ejected baryons 
provided (in the timestep $\Delta t$) by supernova feedback. 
\begin{eqnarray}
k\,T_*(M')=&{m_H\over 3\,\Delta m_h}\, 
E_{SN}\eta_{SN}\dot m_*\,\Delta t \nonumber~~~~~~~~ \\
=&{m_H\over 3}E_{SN}\eta_{SN}
\big({v_h\over v_c}\Big)^{\alpha_h},~~~~~~~
\end{eqnarray}
where $E_{SN}=10^{51}$ erg/s is the energy per supernova, $\eta_{SN}$ is the 
 number of supernovae per solar mass ($3.2\,10^{-3}$ for the Scalo IMF we 
shall use here); 
the latter equality holds after  point d in Sect. 2.1. Note that 
the ratio $\dot m_*\Delta t/\Delta m_h = (v_h/v_c)^{\alpha_h}$ depends only 
on the current circular velocity of the halo and not on the progenitor masses.  

Finally, the cold gas 
fraction $f_*$ expelled outside the virial radius of $M'$ 
and heated at $T_*(M')$ is computed adopting the Cole et al. (1994) 
recipe, i.e., assuming that the reheated gas is expelled from the halo. 
So the fraction of gas  reheated in the timestep $\Delta t$ reads
\begin{equation}
\Delta f_*= \Delta m_h/m_h~.
\end{equation}
Note that the quantity $\Delta f_*$ depends on the merging history of the 
halo, as will be described in detail in Sect. 2.3.

Our guideline will be that, since the X-ray luminosity in  eq. (1) depends 
strongly on $f_*$ and $T_*$ which in turn depend on the star formation 
and on feedback, 
a strong {\it correlation} must exist between the X-ray emission and 
the SFR. 

\subsection{ Convolutions over the Merging Histories}

Both the SF and the X-ray emission
depend ultimately on the merging history of the halos. In fact, at each merging
event the gas reservoir of a halo are increased by the gas (cold and hot) 
enclosed in the merging partner, and this changes the processes of cooling,
feedback and SF together; in addition, by the 
sequence of overlapping merging events, the infalling gas compresses
the ICM at the cluster boundary, and sets the X-ray luminosity (eq. 1) through
the factor $G^2$. 

So to obtain the observable quantities -- average value 
and scatter -- associated to a halo, we must convolve 
over all merging histories leading to that mass. In detail, we
adopt the following procedure, essentially an analitical rendition of
the SAM, but complemented with
the proper description of the physics of the X-ray emitting ICM. 

i) We define a time-velocity grid, with grid size $\Delta t=t_0/100$ and 
$\Delta v_c=10$ km/s ($t_0$ is the present cosmic time). 

The number density $N(M,z_i)$ of DM halos with mass $M$
[corresponding to a circular velocity $v_c=(10\,G\,H(z)\,M)^{1/3}$] at the
initial redshift $z_i \sim 10$ is taken from the Press \& Schechter (1974)
expression. Initially, to each halo we associate a galaxy with the
same $v_c$ of the halo (more than one galaxy per halo is 
an exceedingly rare circumstance for $z_i\sim 10$).  For each $v_c$ 
the corresponding mass $M$ and virial temperature $T_v$ are
computed. As said above, the initial gas content associated to all
halos corresponds to the universal baryonic density $\Omega_b$, and
the initial stellar content is taken to be nil.

For each circular velocity $v_c$, 
the baryonic content is divided into {\it stars}, 
{\it cold} gas and {\it hot} gas as described 
 in \S  2.1  under point a).

ii) At the next time step,  we compute the mass transfers: hot $\rightarrow$ 
cold $\rightarrow$ stars $\rightarrow$ hot, as described in 
\S 2.1, point b), c) and d).  

iii) For each circular velocity $v_c$, we compute the further variations of 
the mass of the gas components and of the stars content due to merging of 
DM halos, possibly followed by galaxy coalescence. We also compute the 
 compression factor $G^2$ (eqs. 2) and the corresponding X-ray luminosity 
(eq. 1). 

Since the merging process is stochastic, at each time step and for
each $v_c$ (corresponding to $M$) we compute the probability density
$\partial^2 P(M',t|M,)/\partial M'\partial t$ that a halo of mass $M'$
has merged with a halo of complementary mass $M-M,'$ to yield the
considered $M$ in the time interval $\Delta t$; within 
the standard hierarchical clustering, such a probability is provided
by the EPST.

The average mass of the cold and the hot gas contents in a halo with
circular velocity $v_c$ are updated according to following equation,
analogous to White \& Frenk's (1991): 
\begin{eqnarray}
m_h(v_c,t+\Delta t)=& m_h(v_c,t)+\Delta t\int_0^{M(v_c)}dM' \times 
\nonumber \\
 & {N(M',t)\over N(M,t)}
{\partial^2 P(M',t|M)\over \partial M'\partial t}m_h(M')~.
\end{eqnarray}
This yields the average increment due to the merging together with the halos 
of previous baryon reservoirs.  
An analogous equation holds 
for the baryonic mass in stars $m_*(v_c,t)$ and in the cold phase 
$m_c(v_c,t)$, as well as for the fraction $f_*$ of ejected gas, whose 
 time increment is given by eq. (6). 

As for the X-ray luminosity, this is computed from the compression
factor $G^2(M,M')$ (eq. 2) inserted into eq. (1).  Its average value
is computed by an analogous convolution to the above one:
\begin{eqnarray}
\lefteqn{
L_X(v_c,t+\Delta t)=L_X(v_c,t)+\Delta t\int_0^{M(v_c)}dM'\times }
~~~~~~~~~~~~~~~~~~\nonumber \\
&{N(M',t)\over N(M,t)}\,{\partial^2 P(M',t|M)\over \partial M'\partial t}
\times ~~~~~~~~~~~~~
\nonumber \\
&A\,\Big({m_h\over M}\Big)^2\,G^2(M,M')\,T_v^2(M)\rho^{1/2}(z).
\end{eqnarray}

iv) The values of $m_c$, $m_*$ and $m_h$ are reset for every $v_c$ after 
all their (positive or negative) increments due to cooling, SF, 
supernova feedback and merging have been computed in the steps ii) and iii). 

v) Finally, for each $v_c$ the associated optical luminosity at the 
wavelength $\lambda$ is computed by the convolution 
\begin{equation} 
S_{\lambda} = \int_0^t\,\phi_{\lambda}(t-t')\,m_*(t')\,dt'~,
\end{equation} 
where the spectral energy distribution of luminous stars  
$\phi_{\lambda}(t)$ is taken from a canonical model 
of stellar population synthesis in its 
latest version (Bruzual \& Charlot 1998). 

vi) We increment the current time by $\Delta t$, and 
repeat the whole procedure through steps ii) to v) until the output time 
 is reached. 

We comment that our analytical rendition of the galaxy formation sector is 
technically similar to the formulation by White \& Frenk (1991), but 
{\it differs} in a number of respects: a) we consider here all three
components (stars, cold gas phase, hot gas phase) which are involved
in the SF process, while in the paper by White
\& Frenk (1991) only cold gas and stars where considered; b) we implement
the comprehensive recipe used in SAM for the star
formation and feedback, while in Frenk \& White (1991) a
simplifying assumption of self-regulation had been adopted to obtain a
simple expression for the SFR. To compute the optical 
luminosities we have used here a standard population synthesis model in 
the updated version by Bruzual \& Charlot (1998). On the other hand, 
we plan to add chemical enrichment models in an extension of this work. 

Our main improvement concerns of course the insertion in the old
framework of the novel issues concerning on the X-ray
emitting baryons.

\section{CHOICE OF PARAMETERS}

In fact, our aim is to obtain the X-ray counterpart of different 
SF histories. We consider two {\it extreme} cases A and B, 
both consistent with the present data discussed in the Introduction. 

Model A is characterized by a SFR declining beyond $z\approx 2$, as
was originally suggested by Madau et al. (1996), and as was obtained
from the SAM (Baugh et al. 1998). From the latter authors, we adopt
the set of parameters reproduced 
in the first row of our Table 1, and chosen by them
to yield an acceptable fit to the local galaxy luminosity
function. The resulting SFR we find (see fig. 1a) is peaked at
$z\approx 1.5$ and in fact declines considerably for $z\approx 2$.

The corresponding galaxy luminosity function at $z=0$ is shown in
fig. 1b.  As discussed by Cole et al. (1994, see their Table 1; also 
Baugh et al. 1998), 
both the decline of the SFR at large $z$ and the flat shape of the
galaxy luminosity function are due to the strong supernovae feedback at
small $v_c$ resulting from their set of star formation and feedback
parameters. The same feedback affects strongly the X-ray luminosities 
through the compression factor $G^2$ and the factor $m_h^2$ in eq. 1.

Model B instead is characterized by a SFR flat beyond $z\approx 2$,
see fig. 2a.  The set of parameters leading to this are shown in the
second row of Table 1.  Such choice  
corresponds to a milder feedback even in small halos, consistent with 
recent works by Ferrara \& Tolstoy (1999) and by Martin et al. (1999). This 
yields a rather steep local galaxy luminosity function (fig. 2b) below $L_*$,
a feature considered tenable, if marginally, in view of the data by
Zucca et al. (1997).

In deriving the galaxy luminosity functions we followed Cole et
al. (1994) in normalizing the mass to light ration in terms of
$\Upsilon$, the total mass in stars divided by the mass in luminous
stars with mass $>0.1\,M_{\odot}$. We adopted $\Upsilon=2.7$ for Model
A (the value of the fiducial model by Cole et al. 1994), and
$\Upsilon=2$ for Model B.  

For comparison with the SAMs, 
we adopt thir ``fiducial'' cosmological/cosmogonical context, namely the  
CDM model with $\sigma_8=0.67$,
$\Omega=1$, $\Omega_b=0.06$ and $h=0.5$.  In all cases, the Scalo IMF
has been adopted, as widely used in SAMs. 

\section{RESULTS}

The outcomes in X-rays from Model A are shown in fig. 3. 
The $L_X-T$ relation provides a good fit to the data ({\it panel a}) 
down to the sub-keV range, the region most affected by supernova
 feedback (eq. 6). Note the mild evolution out to  $z=1$. 
The corresponding X-ray luminosity function ({\it panel b}) also 
agrees well with the local data (Ebeling et al. 1997; De Grandi et al. 1998) 
and shows little evolution out to $z\approx 1$ 
in agreement with recent data (Rosati et al. 1998). We also plot 
({\it panel c}) the complementary contribution to the soft X-ray background 
(details are given in CMT98a), which is well below the observational limits even when 
a 75 \% resolved contribution is subtracted out of the ROSAT data 
(see Hasinger et al. 1998; Giommi, Fiore \& Perri 1998). 

The corresponding results for the opposite extreme Model B are shown in
fig. 4. Here the $L_X-T$ relation is flatter at low $T$, and constitutes a 
marginal fit to the existing data. 
This is because the lower amount of gas 
expelled by virtue of the smaller feedback
(with a minor effect from the increased temperature $T_*$) 
reduces the changes with $T$ of 
the compression factor $G$ entering the X-ray luminosity
(eq. 1); the result is closer to the gravitational self-similar scaling 
$L\propto T_v^2$. This goes back to the 
circumstance (see fig. 2) that, in the absence of an appreciable 
contribution $f_*\,g^2(T_*)$ from 
supernovae in the merging partner $M'$, the prevailing term 
$(1-f_*)\,g^2(T_v)$
is determined only by the virial temperature ratio $T_v(M)/T_v(M')$. 
Such ratio of purely 
gravitational temperatures does not change appreciably with $T$ when 
averaged with the self-similar merging rates; so a nearly 
constant $G^2$ obtains, and eq. (1) tends to $L\propto T_v^2$. 

Another feature of Model B is the increase with $z$ of the
X-ray luminosities; this is due to the larger amount of hot gas [the factor
$(m_h^2/M^2)$ in eq. 1] made available by the stronger global, supernova
activity, 
but retained in shallow wells by virtue of the weak feedback. 
This has a number of implications: 
not only the faster evolution of the $L_X-T$ relation represented by the dashed 
curve in {\it panel a}, but also the increased number of faint sources
represented in the luminosity functions 
$N(L,z)$ in {\it panel b}; in addition, the contribution
to the soft X-ray background is larger, and barely consistent with the
observational limits. 

But the key X-ray test telling apart the two models is provided by 
the source counts $N(>F)$ shown in fig. 5. The larger number of sources
predicted at high $z$ by model B implies faint counts larger by factors 
$\sim$ 4 relative to Model A; their 
redshift distributions are shown in fig. 6A and 6B.
Correspondingly, in fig. 7 we show for the two models 
the luminosity density in X-rays, 
the counterpart for the hot baryons of the SFR associated with 
condensed baryons  shown in fig. 1. Note that for model B the 
peak is higher and 
shifted to $z\approx 2$, corresponding to the larger SF activity at 
 high $z$. We also stress that, in any case, the effect of SF on the 
X-ray emission is always delayed by the few dynamical times taken by the merging 
 activity (see eq. 8) to affect the ICM. 

Finally, we show in fig. 8 the correlation between the X-ray flux 
and the optical ($B$ and $K$) magnitudes in DM halos, as  
predicted by Model A and by Model B. 

\section{CONCLUSIONS}

In this paper we have related two quantities: the X-ray emission
from the hot diffuse baryons in hierarchically 
evolving potential wells, and the history
of the baryons condensed into stars.  We have presented a computational
package which grafts on the extended Press \& Schechter
treatement of the hierarchical clustering both  
our model for the thermo- and the hydrodynamics
of the X-ray emitting ICM, and our analytical
implementation of the current recipes adopted in the
semi-analytical models of galaxy formation.

We have shown that supernova feedback is an essential
ingredient in deriving not only the early SFR, 
but also 
a realistic shape of the $L_X-T$ correlation for groups and clusters of 
galaxies, 
and in predicting counts and luminosity functions of faint X-ray sources.  
In fact, the SFR at $z\gtrsim 2$ 
depends strongly on the amount and the temperature of the baryons retained is 
{\it small} potential wells. But as such small halos are included into 
larger structures at lower $z$, the same quantities affect 
the X-ray properties of the accreting halos. 

Specifically, we find that if the SFR is peaked at $z\approx 1.5$ with a {\it
decline} to higher $z$, then the local $L_X-T$ correlation 
{\it flattens} strongly going from groups to rich clusters of galaxies, and
evolves little with $z$; the corresponding luminosity density in
X-rays is peaked at $z\sim 1$. Conversely, if the SFR was already
high since $z\approx 4$, then a {\it smoother} local $L_X-T$ relation
obtains, closer to the self-similar form $L_X\propto T^2$ down to poor
groups; relatedly, the counts of such sources brighter than $F_X\simeq 10^{-15}$
erg/cm$^2\, $s exceeed the former case 
by a factor $\approx 4$ in the energy band $\gtrsim 0.5$.  
Even fainter fluxes are within the reach of
the next generation X-ray observatories like AXAF (and XMM, though
near its confusion limit). To be conservative with surface brightness, we 
have included in the counts only $L_X\geq 10^{43}$ erg/s. 
In addition, the above excess 
counts has been computed with a
very conservative low energy threshold (0.5 keV) for AXAF, which in fact 
may have an effective threshold of 0.25 keV (R. 
Giacconi, private communication), while 0.1 keV 
is planned for XMM; for example, on lowering the threshold down to 0.25 keV  
the excess would be doubled (see fig. 5). 

All that will make in the near future the X-ray observations of groups and clusters a 
powerful means to discriminate the strength of the feedback processes, 
which determine also the SF history at high $z$. 
For example, the excess X-ray counts by a factor $4$ corresponds to a 
filling by stellar and hot baryons of the early potential 
wells which yields a SFR larger by $\sim 10$ at $z\approx 3-4$. 

We have investigated the sensitivity of our results to changes of 
the cosmological framework. We find that the 
difference between Model A and B persists when one considers different 
cosmologies within  the usual constraints provided, e.g., 
by the cosmic age and the local counts of X-ray clusters. In particular, 
we have checked that for a flat
$\Lambda$-cosmology ($\Omega_o=0.3$, $\Omega_{\lambda}=0.7$, $\Omega_b=0.04$) 
the count excess is still larger than 3.5. A more extended study of the 
SFR-X-ray connection for different cosmologies will be presented elsewhere. 

We have explored the effect of varying the star 
formation and feedback parameters, and of changing the 
star formation recipe altogether. 
We have first kept $\alpha_*$ fixed at 0, and tried values of $\alpha_h$ 
intermediate between Model A and B, yielding intermediate decline 
of the SFR at $z\gtrsim 2$. We find X-ray counts again exceeding 
those of Model A for any value of $2\leq \alpha_h< 5.5$; 
e.g., at $F_X=10^{-15}$ erg/cm$^2$ s 
for $\alpha_h=3.5$ the excess over Model A is a factor
2.3, while for $\alpha_h=4$ the excess is still factor 1.9; recall that
when the energy band is extended down to 0.25 keV the difference between
this case and Model A is doubled. Thus, X-ray
observations with low energy threshold, as planned for the next space 
observatories, can pinpoint the effective
strength of the feedback within the range delimited by our extreme 
Models A and B. If we now change $\alpha_*$ to the other extreme value 
-1.5 the above results are not changed appreciably. This is because 
it is the feedback which dominates (see Cole et al. 1994) the amount and 
the thermal state of 
all baryons in shallow, early  potential wells, and so governs both the 
X-ray emission and the SFR at large-$z$. 

We have also investigated the effect of adopting recipes for star
formation and for feedback different from Cole et al. (1994).  
In particular, we implemented in detail the recipe by
Kauffman et al. (1998), where all the gas is retained in the halos and
the feedback parameter has about the same $v_c$-dependence of our Model B. 
This yields a flat SFR at $z>2$;  but in addition we obtain X-ray counts
close to those in Model B, confirming the trend: more stars - more
X-rays.  However, the lower normalization of the feedback leads to a very flat
local $L_X-T$ correlation.

Finally, we have checked that our results  are not sensitive to the inclusion of
additional heating or cooling. Specifically, the 
supernova heating of the hot fraction inside the deep wells turns out to be 
not relevant, unless the associated $T_*$
is forced up to values close to the virial temperature 
around $10$ keV,  at variance with the
accepted star and supernova energetics. Such findings are consistent
with those by Metzler \& Evrard (1997) based on hydrodynamical/N-body
simulations. On the other hand, considering explicitly the cooling of the 
gas  ejected beyond the virial $R$, the temperature $T_*$ decreases 
by only $5-10 \%$, due
to the low densities of the gas in the external regions. This holds both for a
distribution spread outside $R$ with a density following the
DM's, and for a constant density shell of thickness $v_{esc}\,\Delta 
t$ located at $R$.  

In sum, the thermal state and amount of hot baryons (with 
their X-ray emission) and of the condensed baryons in shallow potential wells 
(yielding the SFR at high $z$ observed in the O-UV bands) both depend 
on the same energy feedback from supernovae. The X-ray and the O-UV  bands 
concur to yield predictions and information concerning the baryon history; 
in particular, the X-rays catch {\it directly} the feedback in the act and 
probe its effective strength. 
We add that, while in this paper we have focussed on the truly 
hot component (at $T \sim 10^6-10^8$ K, making up a 
minor but vocal fraction of all baryons), the connection here 
investigated involves other baryonic phases, observable in different bands; 
e.g., the lukewarm ($T<5\,10^4$ K) gas contributing to the 
the different SF histories can be compared with 
the amount of photoionized gas probed by absorbtion in the Ly-$\alpha$  
clouds (both in and out the galaxies). 
Such a variety of independent probes will stricly bound the amount and 
behaviour of the feedback processes, 
which are presently highly uncertain but in fact 
govern the early cosmic SFR  and so 
the emission properties of the faint galaxy population. 

Acknowledgements: this work has been triggered 
by a discussion with R. Giacconi 
who posed the question of any connection between the planned 
AXAF deep counts and the optical SFR. Informative  
discussions with P. Rosati are 
also gratefully acknowledged. We thank our referee for his stimulating 
and helpful comments.

\newpage
\section*{FIGURE CAPTIONS}
\figcaption[]{ 
Panel a): The global star formation rate, as a  function of redshift $z$ for 
Model A. 
\hfill\break
Panel b): The B-band luminosity function of galaxies in Model A; for 
comparison, we show the data by Zucca et al. (1998). 
\label{fig1}}

\figcaption[]{ 
Same as fig. 1, but form Model B.
\label{fig2}}

\figcaption[]{ Results from Model A:\hfill\break  
Panel a):  
The $L_X-T$ correlation at $z=0$ (solid line) and at z=1 (dashed line). 
Group data from Ponman et al. (1996, solid squares); cluster data from 
  Markevitch (1998, open triangles). 
\hfill\break
Panel b): The local (solid line) and the $z=1$ (dashed line) luminosity 
function. For comparison we show the data by Ebeling et al. (1997). 
\hfill\break
Panel c): The contribution of hot baryons to the soft X-ray background 
for sources with $F_X< 4 10^{-14}$ erg/cm$^2$s, is compared  with the 
total observed values (open stars, Hasinger et al. 
1997), and with the 25\% unresolved component 
(see Giommi, Fiore, \& Perri 1998, solid squares). 
\label{fig3}}

\figcaption[]{
Same as fig. 2, but for Model B
\label{fig4}}

\figcaption[]{
The source counts from Model A (solid line) and from 
Model B (dashed line), in the energy band 0.5-2 keV. 
The dashed region corresponds to the 
observed ROSAT cluster number counts by Rosati et al. (1998).  
\label{fig5}}

\figcaption[]{
The redshift distribution of X-ray sources is plotted 
 for differents fluxes: 
$10^{-15}< F < 10^{-14}$ erg/cm$^2$ s (solid line), and the very faint range  
$10^{-16} < F < 10^{-15}$ erg/cm$^2$ s 
(dashed line). 
Panel A: Model A; panel B: Model B. 
\label{fig6}}

\figcaption[]{
The X-ray luminosity density (for photon energies in the range 
 0.1-10 keV) as a function of redshift. 
Panel A: model A; panel B: model B. These constitute the X-ray 
counterparts of the O-UV luminosity density corresponding to the 
SFR shown in fig. 1
\label{fig7}}

\figcaption[]{
The correlation between X-ray fluxes and optical magnitudes is plotted 
(shaded region) for  $B$ magnitudes (upper two panels) and 
$K$-magnitudes (lower two panels), 
 for Model A (left column) and Model B (right column)
\label{fig8}}

\large
\begin{deluxetable}{ccccc}
\tablewidth{420pt}
\tablenum{1}
\tablecaption{Model Parameters}
\tablehead{
\colhead{} &
\colhead{$\tau_*^o$} &
\colhead{$\alpha_*$} &
\colhead{$v_{h}$} &
\colhead{$\alpha_{h}$} 
}

\startdata Model A & 2.8 Gyr & -1.5 & 140 km/s & 5.5 \nl & & & &
\nl Model B &  2.8 Gyr & 0 & 140 km/s & 1.5 \nl \enddata
\tablenotetext{}{ The parameters corresponding to the two reference
SFRs introduced in the text. These define, 
as a function of the halo circular velocity $v_c$, the star
formation time scale $\tau_*(v_c)=\tau_*^0\,(v_c/300~{\rm
km/s})^{\alpha_*}$ and the mass $\Delta m_h
=(v_c/v_h)^{-\alpha_h} \,\Delta m_*(v_c,t)$ reheated by SN feedback.}

\end{deluxetable}


\begin{references}

\reference{}Allen, S.M, \& Fabian, A.C. 1998, MNRAS, 297, L57

\reference{} Balogh, M.L., Babul, A., \& Patton, D.R. 1998, preprint 
[astro/ph9809159]

\reference{}Barger, A., Cowie, L.L., Sanders, D.B., \& Fulton, E. 1999, 
 Nature, 394, 248

\reference{}Baugh, C.M., Cole, S., Frenk, C.S., \& C.G. Lacey 1998, ApJ, 
 498, 504 

\reference{}Bond, J.R., Cole, S., Efstathiou, G., \& Kaiser, N. 1991, ApJ
379, 440

\reference{}Bower, R. 1991, MNRAS, 248, 332

\reference{}Bruzual, A., \& Charlot, S. 1998, in preparation

\reference{}Cavaliere, A., Menci, N., Tozzi, P., 1997, ApJL 484, 1 (CMT97)

\reference{}Cavaliere, A., Menci, N., Tozzi, P., 1998, ApJ, 501, 493 (CMT98a) 

\reference{}Cavaliere, A., Menci, N., Tozzi, P., 1998, preprint [atro-ph9810498] 
(CMT98b)  

\reference{}Cole, S., Aragon-Salamanca, A., Frenk, C.S., Navarro, J.F., \& 
 Zepf, S.E. 1994, MNRAS, 271, 781

\reference{}Connolly, A.J., Szalay, A.S., Dickinson, M.E., Subbarao, M.U., 
 \& Brummer, R.J., 1997, ApJ, 486, L11

\reference{}De Grandi, S., et al. 1998, preprint [astro-ph/9812423]

\reference{}Dickinson, M. 1997, `Color-selected galaxies 
and the HDF', in `Hubble Deep Field' eds. Livio  M. et al., CUP, 219

\reference{}Ebeling, H., Edge, A.C., Fabian, A.C., Allen, S.W., 
Crawford, C.S., \& B\"ohringer, H. 1997, preprint [astro-ph/9701179]

\reference{}Edge, A.C., \& Stewart, G.C. 1991, \mnras, 252, 428

\reference{}Ellis, R. 1998, Nature, 395, 3

\reference{}Ferrara, A. \& Tolstoy, E., 1999, preprint [astro-ph/9905280]

\reference{}Fukugita, M., Hogan, C.J., \& Peebles, P.J.E. 1997, preprint 
 [astro-ph/9712020] 

\reference{}Giommi, P., Fiore, F., \& Perri, M. 
1998, preprint [astro-ph/9812305]

\reference{}Hasinger, G. 1996, A\&AS, 120, 607

\reference{}Hughes, D. et al. 1998, Nature, 394, 241 

\reference{}Kaiser, N. 1986, \mnras, 222, 323

\reference{}Kauffmann, G., White, S.D.M., , and Guiderdoni, B. 1993, MNRAS, 
  264, 201

\reference{} Hasinger, G. et al. 1998, A\&A, 329, 482

\reference{}Lacey, C., \& Cole, S. 1993, \mnras, 262, 627

\reference{}Madau, P., Ferguson, H.C., Dickinson, M.E., Giavalisco, M., 
Steidel, C.C., \& Fruchter, A. 1996, MNRAS, 283, 1388

\reference{}Madau, P., Pozzetti, L., \& Dickinson, M.E.
 1998, ApJ, 498, 106

\reference{}Markevitch, M. 1998, ApJ, 504, 27

\reference{}Martin, C.L., 1999, ApJ, 513, 156

\reference{}Metzler, C.A., \& Evrard, A.E. 1997, prprint [astro-ph/9710324]

\reference{}Meurer, G. Heckman, T.M., 
Lehnert, M.D., Leitherer,  C., \& Lowenthal, J. 1997, AJ, 111, 54

\reference{}Navarro, J.F., Frenk, C.S., \& White, S.D.M. 1997, 
  \apj, 490, 493

\reference{}Peebles, P.J.E. 1993, 
{\it Principles of Physical Cosmology} (Princeton: Princeton Univ. Press)  

\reference{}Pettini, M. et al. 1997, `The Spectra of Star Forming 
Galaxies at High Redshift', in `The Ultraviolet Universe at 
Low and High Redshift: Probing the Process of Galaxy Evolution', 
ed. Waller W. et al., AIP Conf. Proc. 408, 279

\reference{}Poli, F., Giallongo, E., Menci, N., Cristiani, S., \& 
D'Odorico, S. 1999, preprint

\reference{}Ponman, T.J., Bourner, P.D.J., Ebeling, H., \& B\"ohringer, H.
 1996, MNRAS, 283, 690

\reference{}Ponman, T.J., Cannon, D.B., \& Navarro, J.F., 1999, Nature, 397, 
135

\reference{}Press, W.H., \& Schechter, P. 1974, \apj, 187, 425

\reference{} Rosati, P., Della Ceca, R., Norman, C., \& Giacconi, R. 
1998, ApJ, 492, 21

\reference{}Smail, I., Ivison, R.G., \& Blain, A.W. 1997, ApJ, 490, L5

\reference{}Somerville, R.S., \& Primack, J.R. 1998, 
preprint [astro-ph/9802269]  

\reference{}Steidel, C.C., Adelberger, K.L., Giavalisco, M., Dickinson, M., 
\& Pettini, M. 1998, preprint [astro-ph/9811399]


\reference{}White, D.A., \& Fabian, A.C. 1995, MNRAS, 273, 72

\reference{}White, S.D.M., Navarro, J.F., Frenk, C.S., 
\& Evrard, A.E. 1993, Nature, 366, 429

\reference{}White, S.D.M.,  \& Frenk, C.S. 1991, ApJ, 379, 52

\reference{}Zucca, E. et al. 1997, A\&A, 326, 477

\end{references}
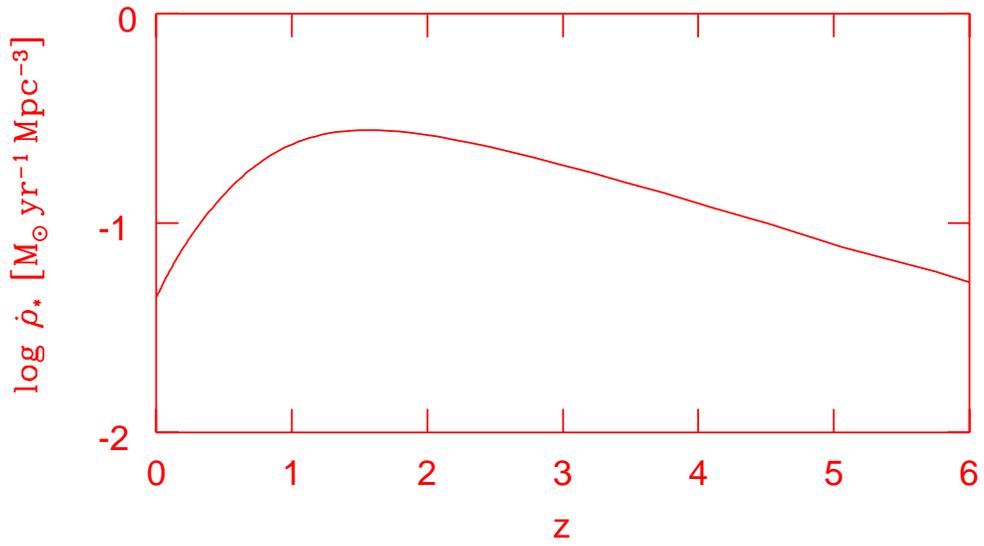
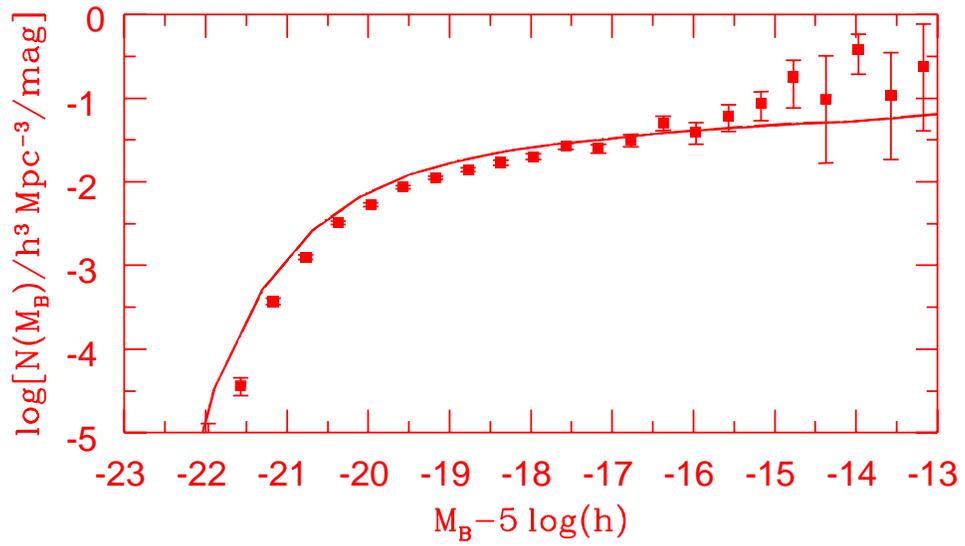
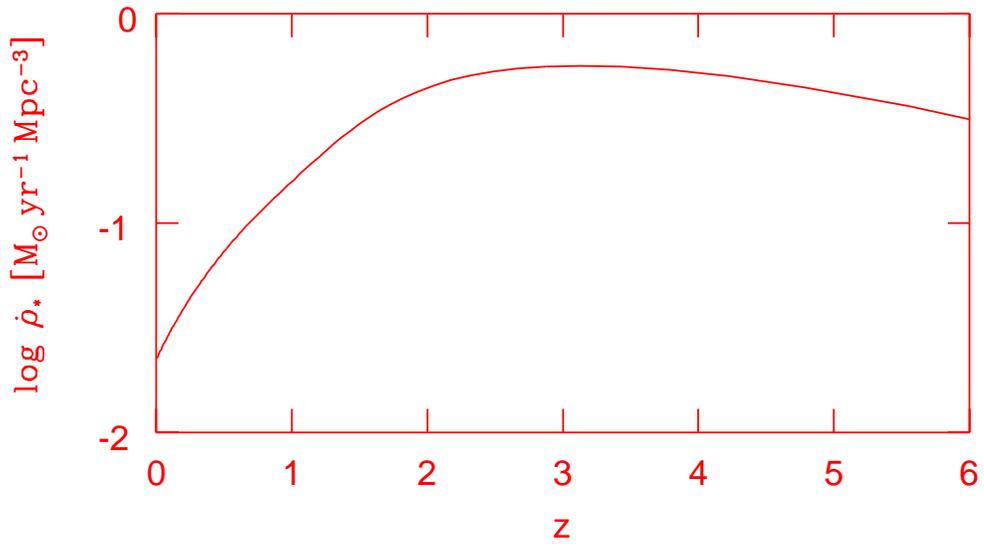
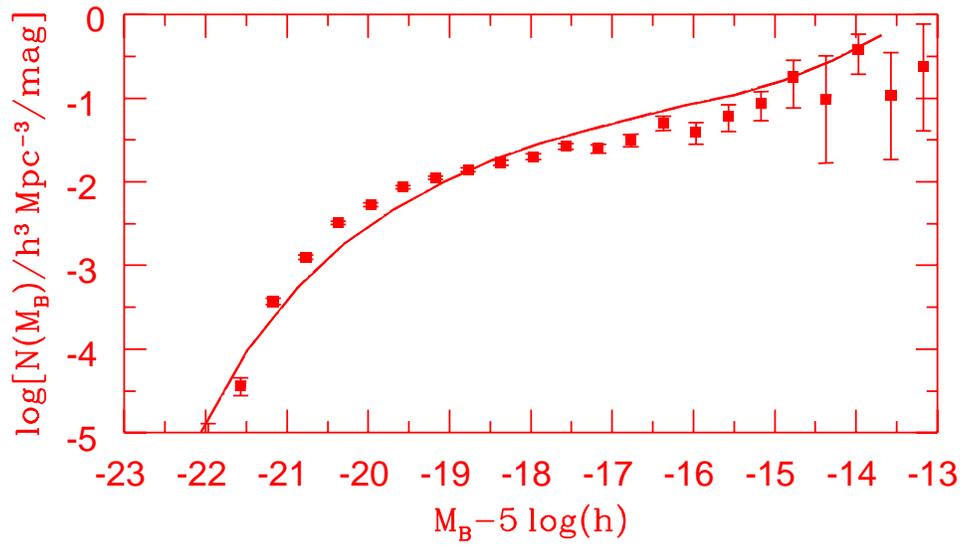
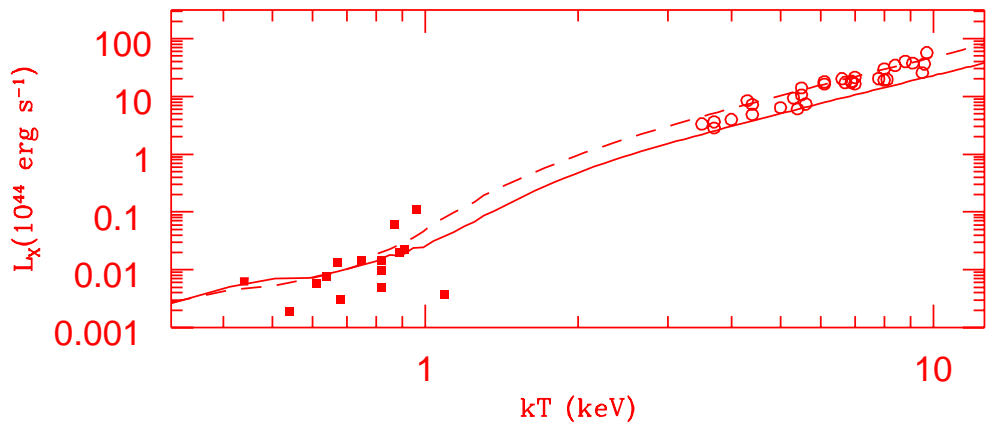
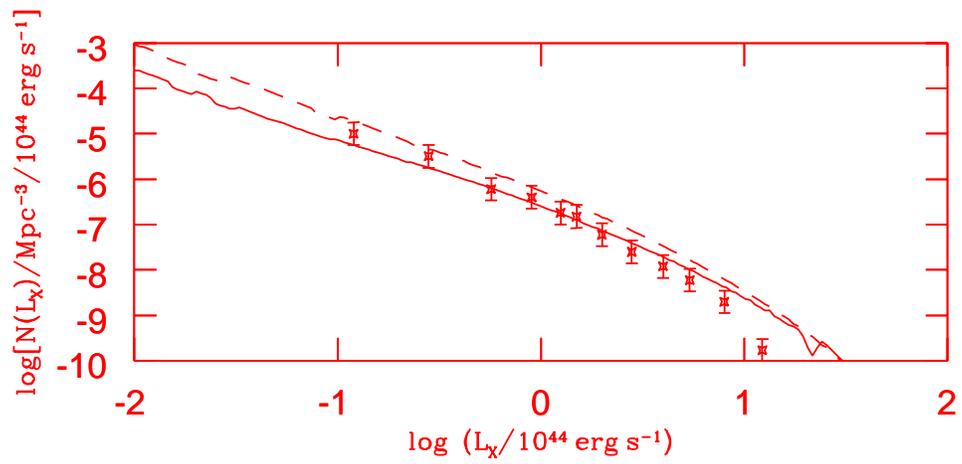
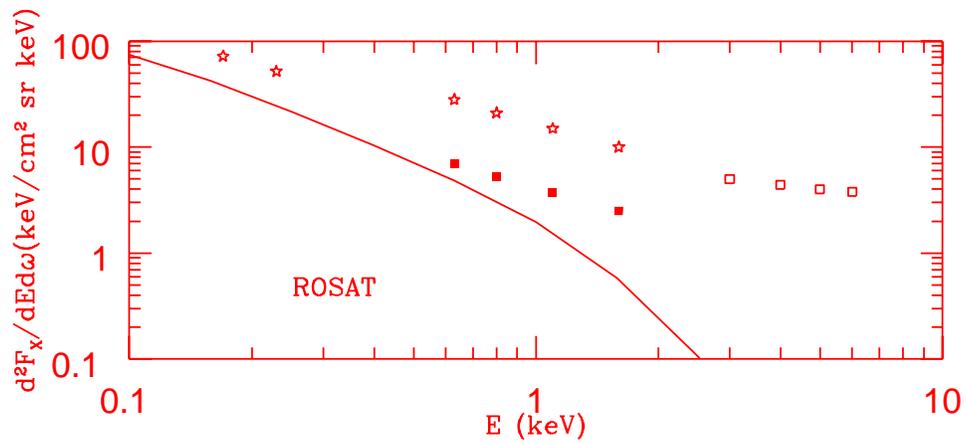
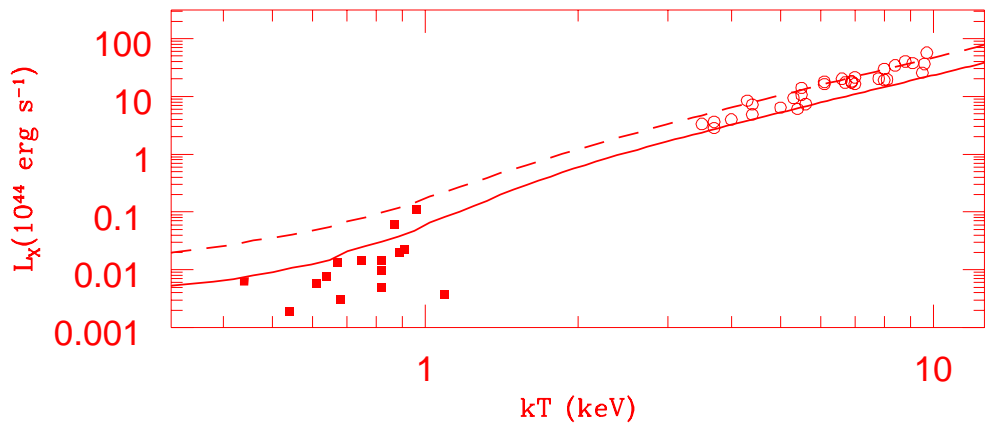
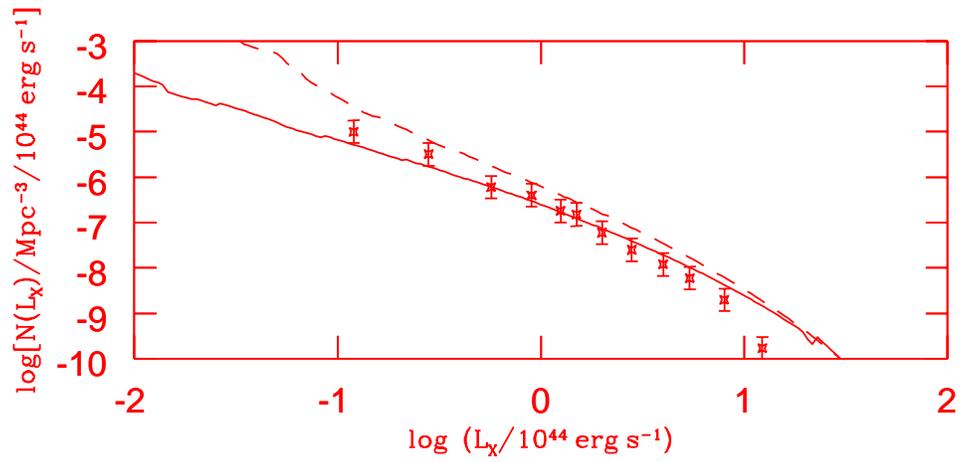
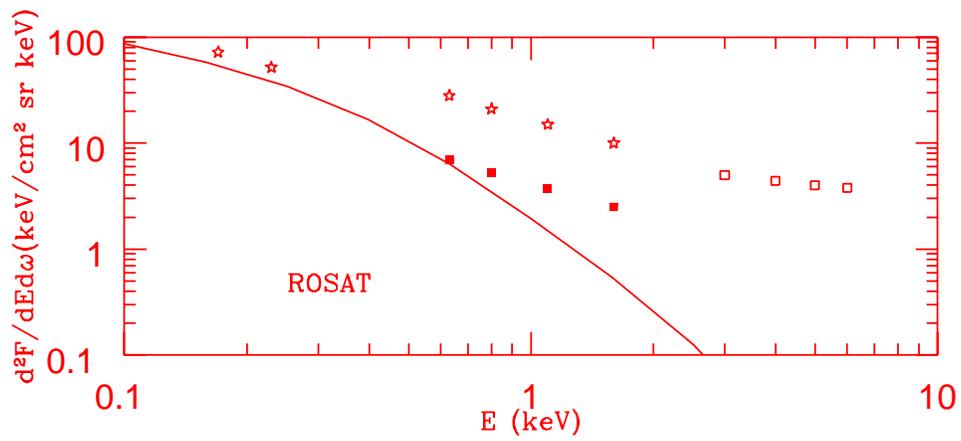
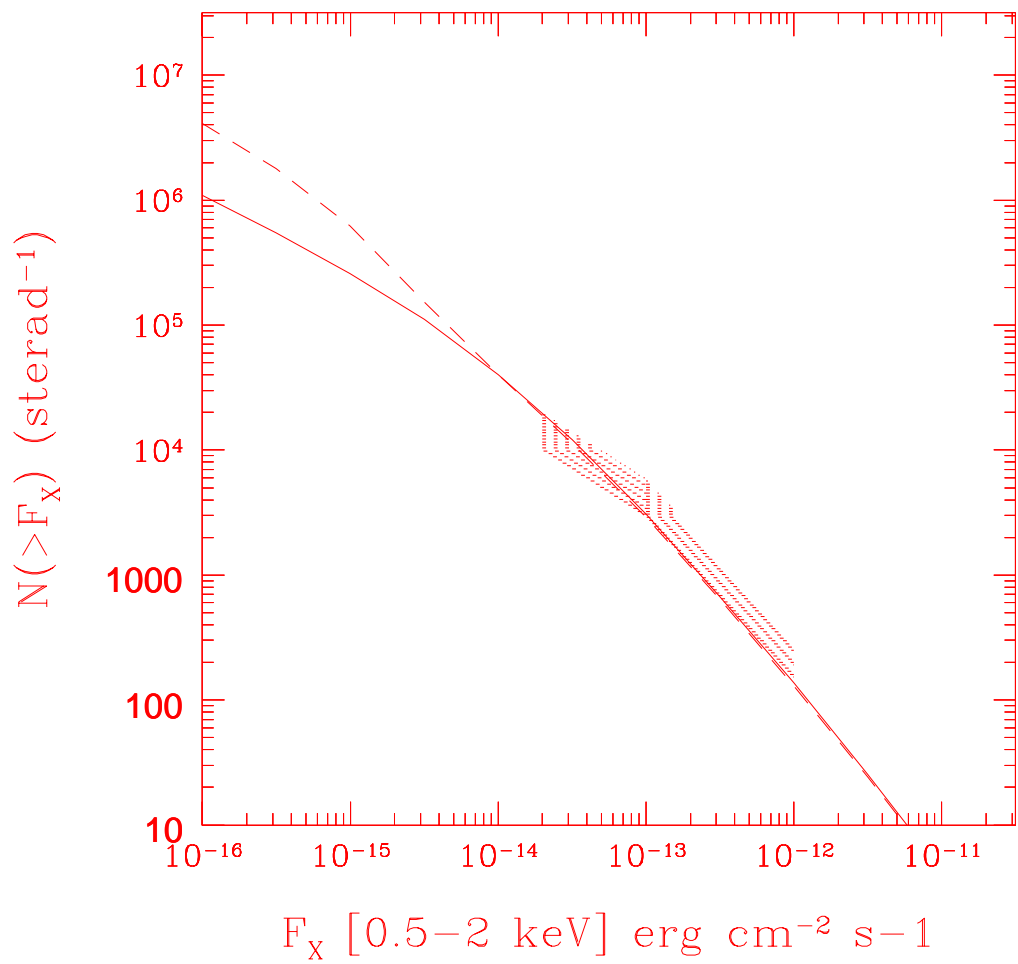
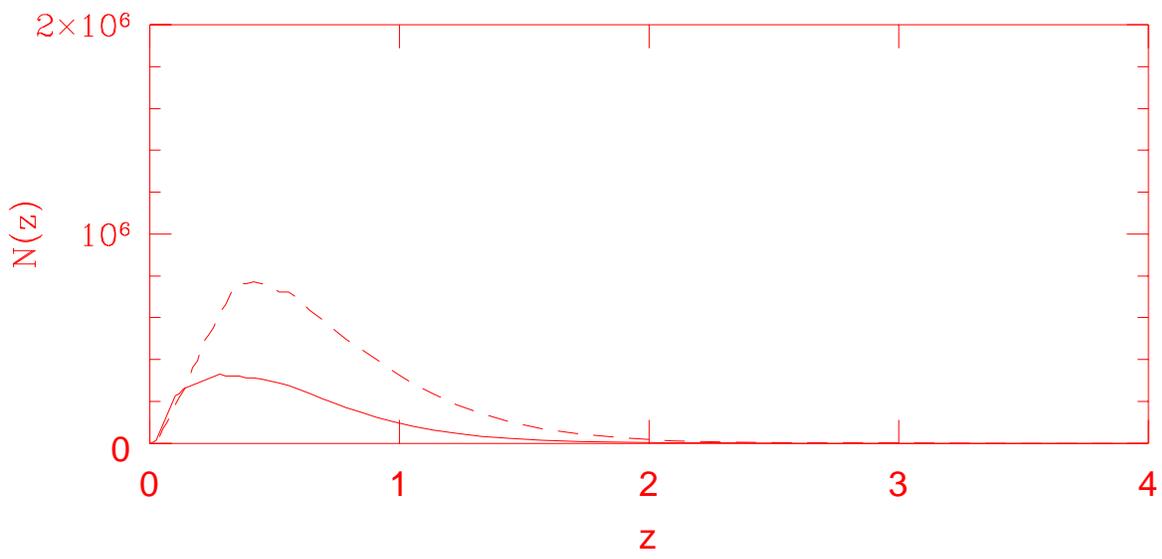
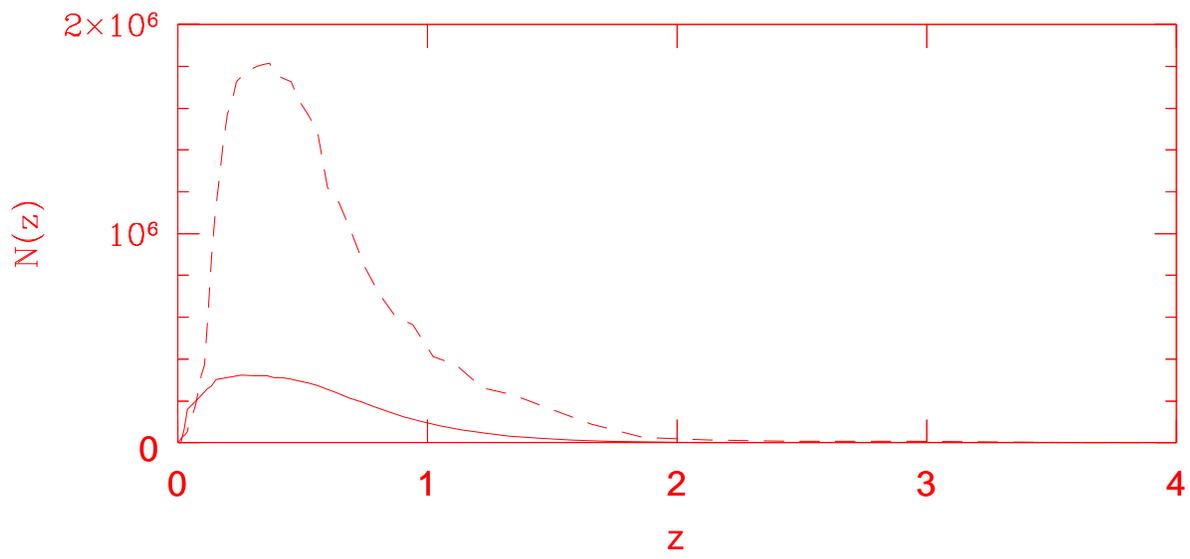
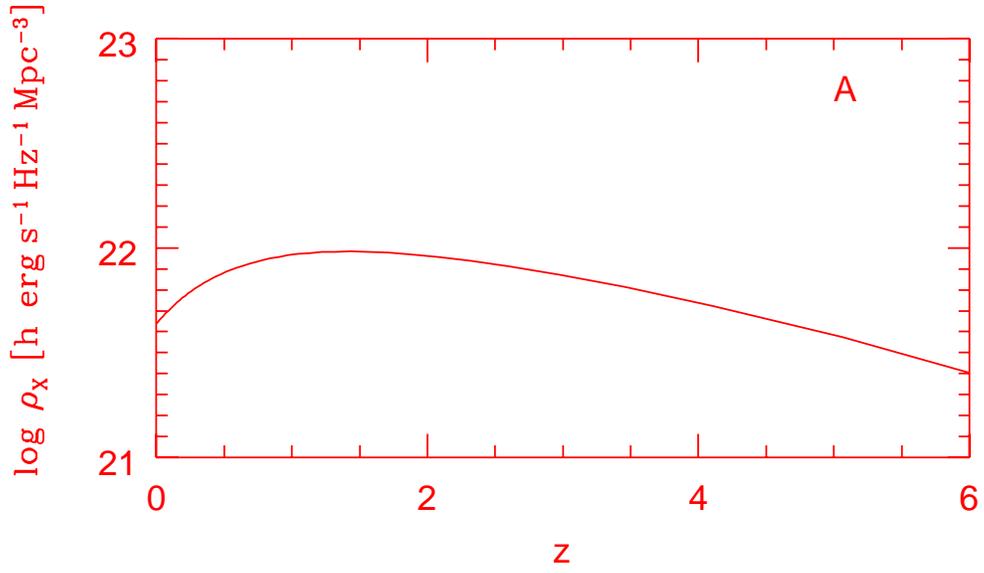
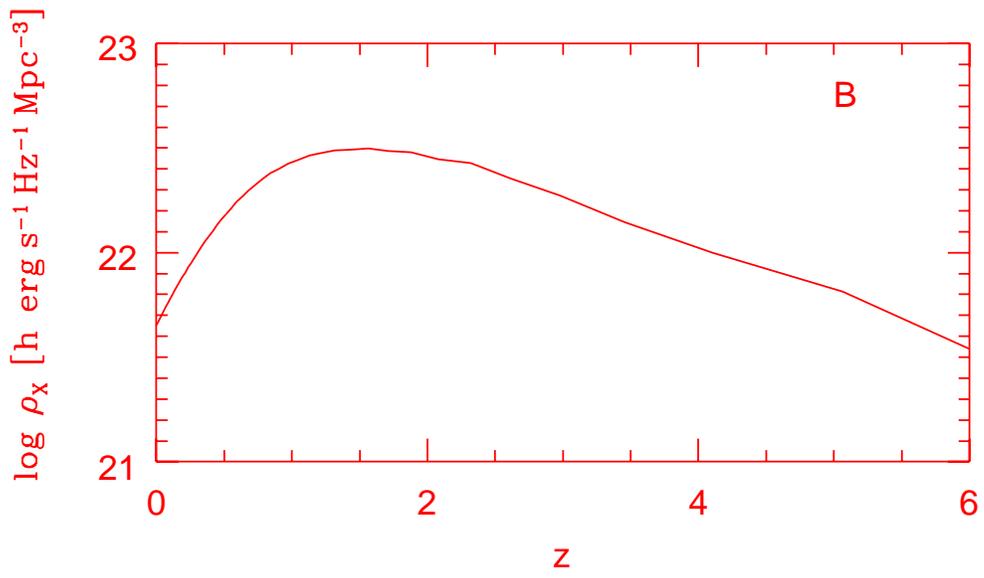
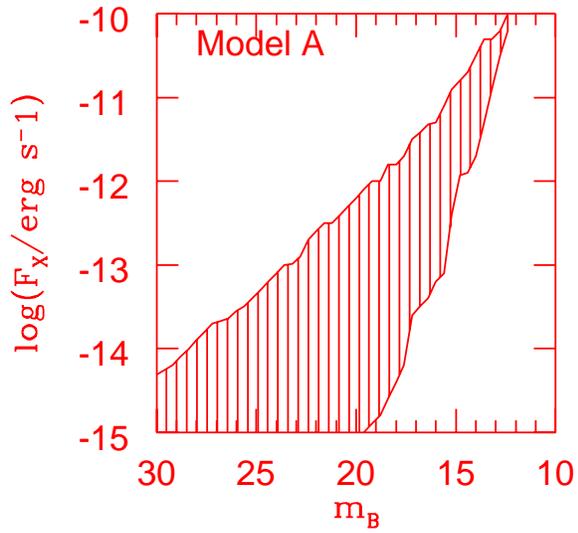
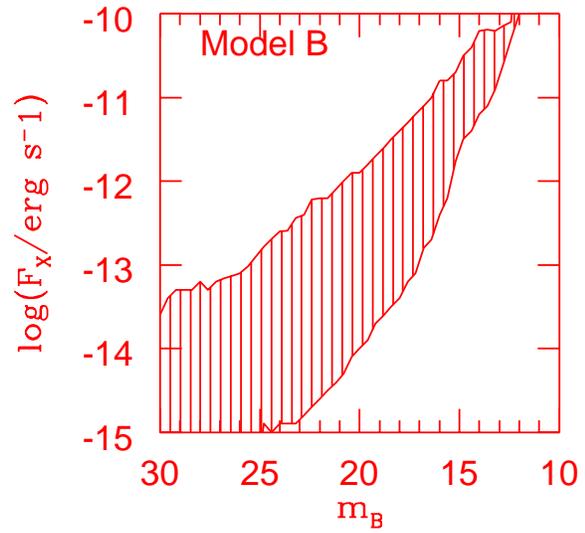
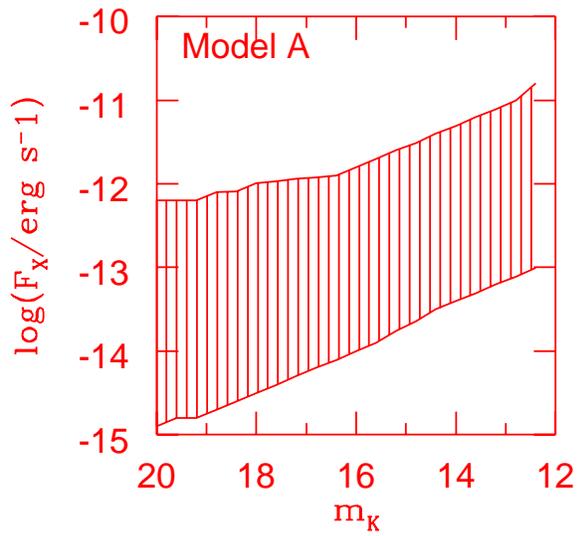
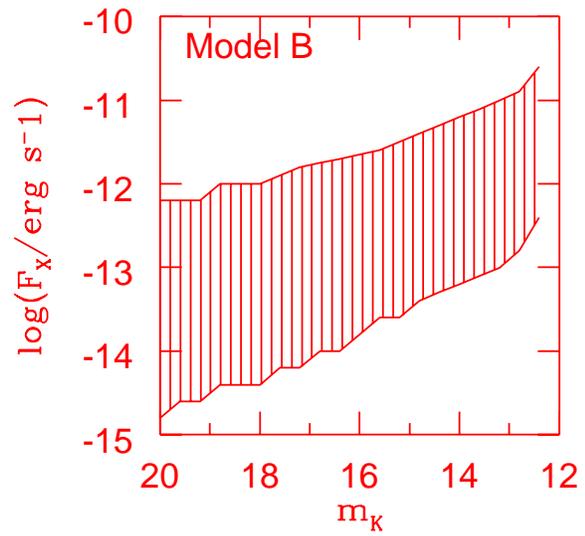
\end{document}